\documentclass[aps,pre,reprint,twocolumn, superscriptaddress]{revtex4-2}
    \makeatletter
\def\@fnsymbol#1{\ensuremath{\ifcase#1\or *\or \mathsection\or \dagger\or \ddagger\or
    \mathparagraph\or \|\or **\or \dagger\dagger
   \or \ddagger\ddagger \else\@ctrerr\fi}}
    \makeatother
\usepackage{bbold}
\usepackage{amsmath,amsfonts,amsmath,mathtools,bbm,bm}
\usepackage{graphicx}
\usepackage{braket}
\usepackage[colorlinks,linkcolor=blue,urlcolor=blue,citecolor=blue]{hyperref}
\usepackage[all]{hypcap}

\usepackage{blindtext}
\usepackage{enumitem}
\usepackage{xcolor}
\usepackage[normalem]{ulem}
\usepackage{float}
\setcitestyle{numbers,square}

\def\({\left (}
\def\){\right )}

\graphicspath{{Figure/PNG/}{Figure/PDF/}{Figure/EPS/}{Figure/TEX/}{Figure/}}

\newcommand{\beq}{\begin{equation}}
\newcommand{\eeq}{\end{equation}}

\newcommand\mydots{\hbox to 1em{.\hss.\hss.}}

\begin{document}

\title{Quenched properties of the Spectral Form Factor}

\author{Dimitrios Charamis}
\affiliation{IPhT, CNRS, CEA, Universit\'e Paris Saclay, 91191 Gif-sur-Yvette, France}
\affiliation{Laboratoire de Physique de l’\'{E}cole Normale Supérieure, ENS, Université PSL, CNRS, Sorbonne Université, Université de Paris, F-75005 Paris, France}

\author{Manas Kulkarni}
\affiliation{International Centre for Theoretical Sciences, Tata Institute of Fundamental Research, Bangalore 560089, India}

\author{Jorge Kurchan}
\affiliation{Laboratoire de Physique de l’\'{E}cole Normale Supérieure, ENS, Université PSL, CNRS, Sorbonne Université, Université de Paris, F-75005 Paris, France}

\author{Laura Foini}
\email{laura.foini@ipht.fr}
\affiliation{IPhT, CNRS, CEA, Universit\'e Paris Saclay, 91191 Gif-sur-Yvette, France}

\date{\today}

\begin{abstract}
{\color{black} The Spectral Form Factor (SFF) is defined for hermitian matrices as the modulus squared of the partition function in complex temperature, with a suitable generalization existing in the non-hermitian case. In this work, we study the properties of what we refer to as the quenched SFF, namely the logarithm of the SFF and in particular its average, and compare them with those of its ordinary (annealed) counterpart, namely the average of the SFF (and eventually its logarithm). 
While the SFF is famously not self-averaging, the opposite is true for the quenched SFF, in both the hermitian and non-hermitian cases. Nonetheless, the quenched and the annealed averages coincide up to subleading constants, at least for high enough temperatures.
The fluctuations of $\ln \mathrm{SFF}$ are deep and one encounters thin spikes when moving close to a zero of the partition function.
In order to study the fluctuations of the quenched SFF at late times we consider a suitable change of variable of $\ln \mathrm{SFF}$ which turns out to be compatible with a Gumbel distribution.
We note that the exponential tail of this distribution can indeed be obtained by sampling the deep spikes of $\ln \mathrm{SFF}$, namely the Fisher zeros of the partition function.
We compare with the results obtained in isolated many-body systems and we show that same results hold at late times also for non-hermitian Hamiltonians and non-hermitian random matrices.
}
\end{abstract}

\maketitle

\section{Introduction}

The spectral form factor (SFF)  is a powerful diagnostic tool used to probe 
 the structure of quantum spectra, capturing both short- and long-range  correlations \cite{Paul_PRL,haake2018quantum,lozej2023spectral}.
Unlike other specific spectral diagnostics, such as the level spacing distribution, the SFF offers time-resolved information that is particularly valuable in systems with chaotic dynamics.

In recent years, the SFF has attracted the attention of different fields of research ranging from quantum gravity (e.g., the Sachdev–Ye–Kitaev model~\cite{sy93,ak1,ak2} and Jackiw–Teitelboim gravity~\cite{saad2019}) 
to condensed matter theory. The renewed interest from high-energy theory - particularly in the context of quantum gravity and holography - was sparked by observations that the late-time behavior of the SFF in models like the SYK model and in ensembles of random Hamiltonians has been used to probe the fine-grained structure of black hole microstates and the onset of quantum chaotic behavior in gravitational systems \cite{cotler2017black,AA_JS}.

At the same time from the condensed matter community the SFF has received considerable attention in the study of random or disordered systems, where it provides insight into the crossover between integrable and chaotic dynamics, as well as between localized and ergodic phases \cite{chan2018spectral}. This non-local measure of spectral correlations and its utility in characterizing and distinguishing between the distinct eigenstate phases of quantum
chaotic and many-body localized systems has been extensively investigated~\cite{PPK2019,Prasad_2024,CH17,L18,GH18,IK1,LS5}. The non-Hermitian generalization of SFF has also gained considerable interest~\cite{fyodorov1997almost,fyodorov1998universality,PhysRevLett.127.170602,GGK22}. More recently, SFF built out of singular values, called $\sigma$FF have been investigated~\cite{sig1,sig2,sig3}. 

A particularly interesting aspect of these studies is the distinction between the SFF computed in a single realization of disorder (the ``sample") and its ensemble average. While ensemble averaging smooths out fluctuations and reveals universal structures—such as the well-known dip-ramp-plateau behavior typical of random matrix theory (RMT)—single-sample SFFs can retain significant non-universal features and exhibit sample-to-sample variability that encodes rich physical information. This distinction has prompted growing interest in understanding the statistical structure of the SFF itself, including questions of typicality, self-averaging, and the emergence of universality in complex quantum systems. 

In this work we study in detail the {\it quenched} SFF, namely the average of the logarithm of the SFF and its fluctuations, comparing with its commonly used {\it annealed} definition {\color{black}(where one considers the average and the fluctuations of the SFF itself)}.
In particular, we consider the distribution of the logarithm of the SFF properly rescaled.
We show that the fluctuations of the quenched SFF are sharp spikes induced by the proximity of a zero of the partition function and that, in a finite time interval, increasing the system size, these fluctuations are suppressed compared to the average. This is in contrast to the fluctuations of the SFF itself, which are of the same strength of its average, the SFF being a non self-averaging quantity.
In doing so we clarify and expand the results of Ref.~\onlinecite{bunin2024fisher}.
We discuss in detail the Gaussian behaviour of the fluctuations of the partition function, which follows from the central limit theorem, in several systems: the Random Energy Model with Poissonian spectrum, a chaotic spin glass at high temperature and its generalisation as a non-hermitian model.
We find that, despite the difference in the fluctuations of the quenched and annealed SFF, their averages coincide in the regime discussed here. Nonetheless the quenched definition would be particularly relevant if one wants to access the low temperature phase of a disordered system.

{\color{black} In Section \ref{SecII} we define the SFF and we describe its phenomenology. In Section \ref{SecIII} we describe the annealed and the quenched average of its logarithm, adressing different issues and in particular its self-averaging behaviour.
Section \ref{SecIV} is devoted to studying the distribution of the logarithm of the (rescaled) SFF at late times.
In Section \ref{Sec_fluct} we discuss the fluctuations of this quantity and in particular how the exponential tail of the distribution can be recovered by making some assumption on the properties of the zeros of the partition function.
Section \ref{SecVI} contains the numerical results for the Random Energy Model which displays Poissonian spectrum, a many-body spin glass in its high temperature phase and its generalization to the non-hermitian case.
In Section \ref{SecVII} we conclude and in
Appendix \ref{Sec_App} we present some results for non-hermitian random matrices. Datasets and numerical code used in this work are openly available on Zenodo \cite{Zenodo}.
}

\section{Ensembles and definitions}\label{SecII}

We will consider the Spectral Form Factor (SFF) for different ensembles of matrices.

\vspace{0.3cm}
In the hermitian case, given a matrix (Hamiltonian) $H$ and its set of ${\mathcal D}$ eigenvalues $\{ E_i \}$, the partition function $Z$ and the SFF are defined as follows:
\beq\label{Def_GOE}
\begin{array}{ll}
\displaystyle
Z(\beta) & \displaystyle = \sum_{i=1}^{{\mathcal D}} e^{-\beta E_i}
\\ \vspace{-0.2cm} \\
\displaystyle
\mathrm{SFF}(t)  & 
\displaystyle
= |Z(\beta + i t)|^2 = |\text{Tr} e^{-(\beta+i t) H}|^2
\\ \vspace{-0.2cm} \\
\displaystyle
& \displaystyle = \sum_{i,j=1}^{\mathcal{D}} e^{- \beta (E_i+E_j) + i t (E_i-E_j)},
\end{array}
\eeq
where we generally introduce a finite inverse temperature $\beta$. At $\beta=0$ the SFF is simply the Fourier transform of the two point correlation function between energy levels $\langle \rho(E)\rho(E')\rangle$.
At finite temperature it can be viewed as the modulus squared of the partition function for complex (inverse) temperature $\beta+i\,t$.
Note that sometimes in the literature different normalizations are chosen but here we chose to identify it with $|Z(\beta+i t)|^2$ without any normalization.
{\color{black} One is typically interested in the average SFF and can furthuer define the connected SFF as
\beq
\label{eq:kc}
\begin{array}{ll}
\displaystyle
\kappa_c(t) &
\displaystyle
= \Big\langle \Big|Z(\beta+ i t)-\langle Z(\beta+ i t)\rangle\Big|^2\Big\rangle
\\ \vspace{-0.2cm} \\
\displaystyle
& 
\displaystyle
= 
 \langle \mathrm{SFF}(t) \rangle - \Big|\langle Z(\beta + i t) \rangle \Big|^2.
\end{array}
\eeq
}
For matrices belonging to the Gaussian Orthogonal Ensemble (GOE) or Gaussian Unitary Ensemble (GUE) it is well known that the average of SFF given in Eq.~\eqref{Def_GOE} shows(see e.g.~\cite{haake2018quantum} {\color{black}or \cite{Forrester_2024_RWalk} and references therein for a detailed formal approach on analytics of the SFF}):
\begin{itemize}
\item A {\it slope} at early times (due to uncorrelated contributions)
\item
A {\it ramp} at intermediate times (associated with the onset of long-range level repulsion)
\item
A {\it plateau} at late times (signaling saturation and the discreteness of the spectrum).
\end{itemize}
As we discuss later, the transition between the early time slope and the ramp at time $t = t_{dip}$ is the point where the fluctuations of $Z(\beta+i t)$ dominate with respect to its average.
For the sake of completeness, we give here the analytical expression for the GUE \cite{haake2018quantum} (see \cite{Br_zin_1997} for the exact expression):
\begin{align}
  \langle\mathrm{SFF}(t)\rangle
  &= \mathcal{D}^2\,\frac{J_1^2(2t)}{t^2}
    + \kappa_c(t), \label{eq:GUE_SFF_a}\\
  \kappa_c(t)
  &= \mathcal{D}
    \begin{cases}
      \dfrac{t}{t_H}, & t \le t_H,\\[1.5ex]
      1,               & t \ge t_H,
    \end{cases}
    \label{eq:GUE_SFF_b}
\end{align}
where $J_1(x)$ is the Bessel function of the first kind of order 1 and $t_H\approx\mathcal{D}\frac{\pi}{2}$.

As we have mentioned, the ramp is a signature of level repulsion and for a Poissonian spectrum (as for the Random Energy Model that we will consider in the following) the ramp is not present.

In the following we will consider also non-hermitian matrices characterised by a set of ${\mathcal D}$ complex eigenvalues $\{\lambda_i = x_i + i y_i\}$.
In this case, it was suggested to define the Dissipative Spectral Form Factor (DSFF) as follows~\cite{fyodorov1997almost,fyodorov1998universality,PhysRevLett.127.170602}
\beq
\label{eq:DSFF}
\begin{array}{ll}
\displaystyle
\mathrm{DSFF}(t_1,t_2)  & \displaystyle
=  | Z(t_1,t_2) |^2
= \Big|\sum_{j=1}^{\mathcal{D}} e^{i t_1 x_j + i t_2 y_j} \Big|^2 
\\ \vspace{-0.2cm} \\
\displaystyle
& \displaystyle 
= \sum_{i,j=1}^{\mathcal{D}} e^{i t_1 (x_i-x_j) + i t_2 (y_i-y_j)}.
\end{array}
\eeq
One could eventually introduce a real temperature also here in an analogous way but its meaning is less clear.
One can nonetheless view this quantity as the squared amplitude of a complex function, depending on two complex variables, where in the above definition we have chosen to look at it in the plane $\beta_1=\beta_2=0$.

If we parametrise $t_1=t \cos\phi$ and $t_2=t\sin\phi$ it turns out that the average SFF for Ginibre Unitary Ensemble (GinUE) depends only on $t = \sqrt{t_1^2+t_2^2}$ and not on $\phi$ \cite{PhysRevLett.127.170602}.
For the Ginibre Orthogonal Ensemble (GinOE), there is some dependence on the angle $\phi$ because complex eigenvalues appear in pairs of complex conjugate numbers.
We will discuss the consequences of this in the following.
For GinUE matrices, one can obtain analytical results, and in particular in the large size limit, one has \cite{PhysRevLett.127.170602}:
\begin{equation}
    \label{eq:DSFF_GinUE}
   \langle\mathrm{DSFF}(t_1,t_2)\rangle=\mathcal{D}+\mathcal{D}^2\frac{4J_1^2(t)}{t^2}-\mathcal{D}e^{-\frac{t^2}{4\mathcal{D}}}.
\end{equation}
This behavior has been compared with the results obtained for spectra of chaotic quantum channels \cite{BOWnARROW_lI_PROSEN_CHAN}, Liouvillian superoperators in open quantum systems (see e.g. \cite{PhysRevLett.127.170602,pawar2025,PhysRevA.110.032220,BOWnARROW_lI_PROSEN_CHAN}), and disordered interacting non-Hermitian Hamiltonians \cite{PhysRevB.106.134202}, such as the Hatano-Nelson model \cite{Hatano-Nelson_OG}.
In the following, for the sake of simplicity, we will consider real matrices that fall into the GOE or GinOE universality class.
For GOE matrices, one observes an almost linear ramp, and explicit formulas can be found in Ref.~\cite{mehta2004random}. Despite the absence of an explicit formula for all complex times parametrized by $(t,\phi)$ for the GinOE (see \cite{cipolloni2023dissipativespectralformfactor} for an expression valid for $t\ll\mathcal{D}^{2/7}$ for real Ensembles), it has been numerically verified that it behaves like that of the GinUE away from $\phi=0,\pi/2$. For the former, the $\mathrm{DSFF}$ is compared to $2\times\mathrm{DSFF_{GinUE}}$ due to the projected degeneracies on the real axis of the complex conjugate eigenvalues. The sub-extensive number of real eigenvalues, which don't come in pairs and scale like $\sim\sqrt{\mathcal{D}}$ \cite{mehta2004random,PhysRevResearch.4.043196, Bordenave_2012, Tao_2015}, preclude an exact match. Near $\phi=\pi/2$, all real eigenvalues are degenerate and hence a deviation from the GinUE value is expected for large $t$. 

\section{The quenched and the annealed SFF}\label{SecIII}

We will be interested in comparing the quenched average of the $\mathrm{SFF}$ (or of the $\mathrm{DSFF}$)
\beq
\label{eq:qu}
 f_q(t) =  \frac{\langle \ln \mathrm{SFF}(t) \rangle}{\ln {\mathcal D}}\, ,
\eeq
and the annealed one:
\beq
\label{eq:an}
 f_a(t)  = \frac{\ln \langle \mathrm{SFF}(t) \rangle}{\ln {\mathcal D}}\, ,
\eeq
where the average is with respect to the ensemble of random matrices and we have normalised in order to obtain a quantity of order one.
In the random matrix case or for physical systems at sufficiently high temperature the ensemble average turns out to be equal to the average over time.

\begin{figure}
    \centering    {\includegraphics[width=8.5cm]{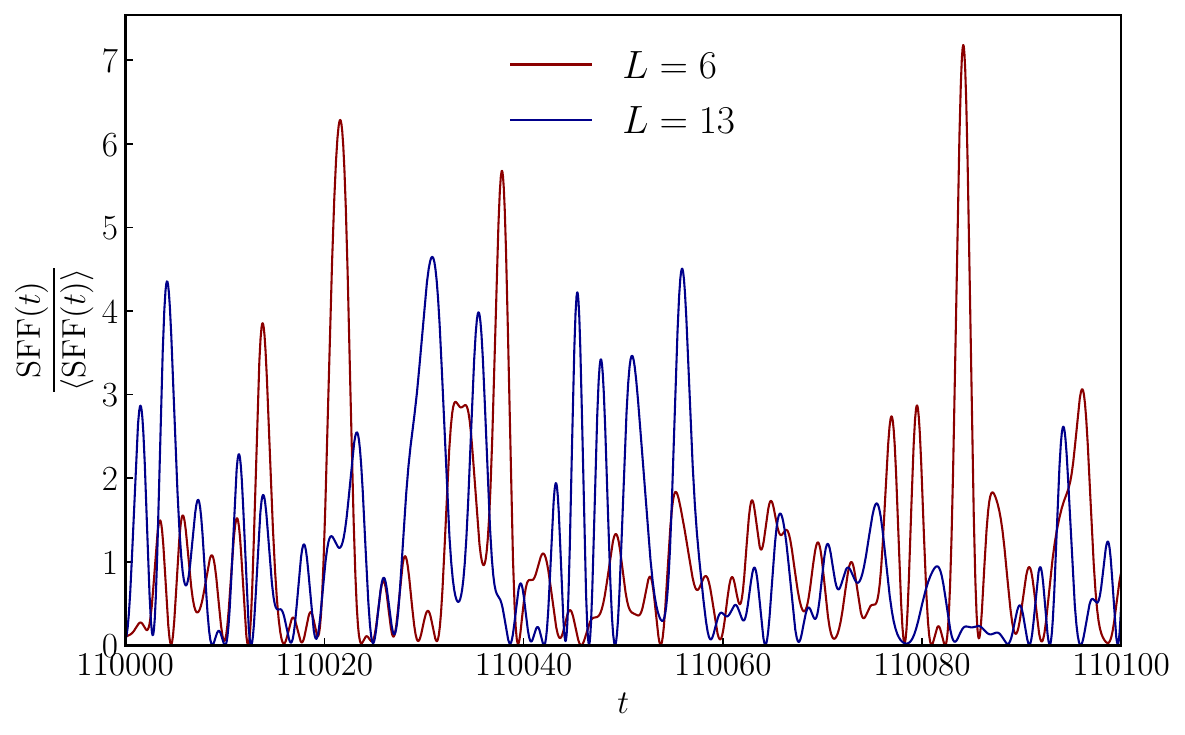}
    \includegraphics[width=8.5cm]{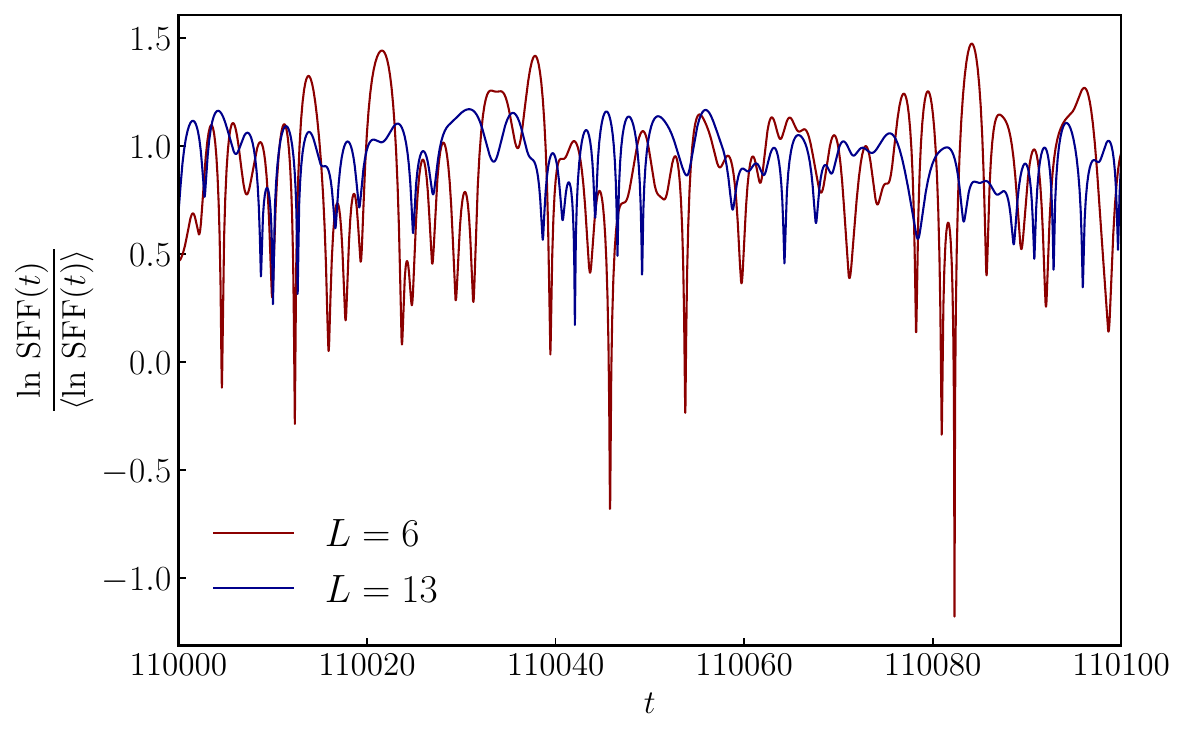}
    }
    \caption{Top: Time sequence of $\mathrm{SFF}(t)/\langle \mathrm{SFF}(t)\rangle$ (upper panel) and $\ln \mathrm{SFF}(t)/\langle \ln  \mathrm{SFF}(t)\rangle$ for the GOE  for two sizes ${\mathcal D}=2^L$ and $L=6$ (red) and $L=13$ (blue) in the plateau regime. The deep spikes of $\ln \mathrm{SFF}(t)$ are induced by the proximity of a zero of $Z(\beta+i t)$, namely, anomalously small values of $\mathrm{SFF}$.
    }
    \label{fig:GOE_PLOT}
\end{figure}

As it has been noted long ago the SFF is not self-averaging \cite{prange1997spectral}. In particular
\beq
\label{eq:fl}
\langle \mathrm{SFF}^2(t)\rangle - \langle \mathrm{SFF}(t)\rangle^2 \simeq \langle \mathrm{SFF}(t)\rangle^2
\eeq
for $t > t_{dip}$ and the fluctuations are of the same order as the average [see Eq.~\eqref{eq:fl}].
In the top panel of Fig.~\ref{fig:GOE_PLOT} we plot the time dependence of an instance of the $\mathrm{SFF}/\langle \mathrm{SFF}\rangle$ for the GOE ensemble for two sizes and we show that the fluctuations do not decrease with the size. Note that the lack of self-averaging in SFF and also in survival probability has been addressed and quantified in different contexts~\cite{LS1,LS2,LS3,LS4}.

The quenched average of the SFF can be computed with the replica method. In fact, it is known in RMT that the SFF behaves as the modulus squared of a complex Gaussian variable $Z(\beta+i t)$ in the ramp/plateau regime. But as we argue,  the Gaussianity of the SFF is more general, i.e., beyond RMT. This assumption is justified for late times and, in many-body systems for sufficiently high temperatures.
Under this assumption its moments are readily computed (see for instance \cite{altland2025statistics} where deviations from Gaussianity have been considered)
\beq
\langle \mathrm{SFF}^n \rangle = n! \langle \mathrm{SFF}\rangle^n + O(\mathcal{D}^{-1}) \ .
\eeq
Note that high order moments have been computed in different many-body systems beyond RMT ~\cite{Flack_2020,Chan_2021,legramandi2025moments} 
.
One can use this result taking the $n\to 0$ limit in the replica method and obtain the quenched value of the $\mathrm{SFF}$
\begin{equation}
    \label{eq:QA}
    f_q(t)\simeq f_{a}(t) -\frac{\gamma}{\ln{\mathcal{D}}}
\end{equation}
with $\gamma$ the Euler gamma constant.
As in the plateau $\langle \mathrm{SFF}\rangle \simeq \mathcal{D}$, this equation shows that {\it quenched = annealed} because up to a subleading constant $\frac{\gamma}{\ln{\mathcal D}}$ the two expressions coincide, despite $\mathrm{SFF}$ being {\it not self-averaging}.
The equivalence between quenched and annealed SFF given in Eqs.~\eqref{eq:qu} and ~\eqref{eq:an} holds away from spin glass phases where it is known to break down, but in this manuscript we will restrain to sufficiently high temperatures.
Nonetheless the possibility to have non-trivial low temperature phases highlights the importance to distinguish between annealed and quenched averages.
 Note that the quenched SFF have been computed by replica method for disordered, periodically driven spin chains in the limit of large local Hilbert space dimension~\cite{Chan_2021}.

The fluctuations of log SFF are the zeros of the SFF which leads to some very deep but very ``thin" spikes \cite{bunin2024fisher}.
These spikes are the thin fluctuations towards small values of $\ln \mathrm{SFF}(t)$ in the lower panel of Fig.~\ref{fig:GOE_PLOT} where we present $\ln \mathrm{SFF}/\langle \ln \mathrm{SFF}\rangle$ for one instance for two difference sizes.
The fluctuations of the quenched average can actually be computed by the replica method and despite the fact that a spike can be very deep the quenched $\mathrm{SFF}$ is self-averaging. In fact one finds:
\beq
\langle (\ln \mathrm{\mathrm{SFF}})^2\rangle - \langle\ln \mathrm{SFF}\rangle^2 \simeq \frac{\pi^2}{6}
\eeq
and the fluctuations are suppressed with respect to the average $\langle\ln \mathrm{SFF}\rangle^2$ as can be seen in Fig.~\ref{fig:GOE_PLOT} comparing two system sizes.
When we look at a fixed time this is true, as the spikes are rare, but when we plot the $\mathrm{SFF}$ as a function of time, for very long time windows which scale as the system's size, we will encounter some spikes and in this sense we will see differences between the single sample and the average which is always smooth.

\section{Change of variable}\label{SecIV}

In order to examine fluctuations of the $\mathrm{SFF}$
we will find useful to consider the rescaled quantity
\beq
\tilde{Z}(\beta+i t) = \frac{Z(\beta + i t)-\langle Z(\beta+ i t)\rangle}{\sqrt{\kappa_c(t)}}
\eeq
and define the variable
\beq
Y= \log |\tilde{Z}(\beta+i t)|^2 \ .
\eeq
The quantity $\kappa_c(t)$ defined in Eq.~\eqref{eq:kc} can be viewed as the connected spectral form factor, which encodes the fluctuations of the $\mathrm{SFF}$ and represents the Fourier transform of the connected two-point function:
\beq
 \langle \rho(E)\rho(E')\rangle - \langle\rho(E)\rangle\langle\rho(E')\rangle \ .
\eeq
{\color{black} It is well known \cite{derrida1991zeroes,Cipolloni_2023} that the crossover time $\tau$ at the end of the slope and the plateau in the Random Energy Model, or the ramp in RMT ($\tau=t_{dip}$), is the point where the fluctuations of $Z$ dominate with respect to its average.}
Therefore, beyond this point one can equivalently consider
\beq
\label{eq:Ydef}
Y= \ln \frac{\mathrm{SFF}(t)}{\langle \mathrm{SFF}(t) \rangle} \qquad \text{for $t \gg \tau$} \ .
\eeq
In RMT, as well as for independent energy levels, it is known \cite{FritzHaake_1999,PhysRevD.98.086026,berezin2020rateconvergencecentrallimit,altland2025statistics}
that $\tilde{Z}(\beta+ i t)$ is a Gaussian complex variable, both in the slope and in the ramp/plateau.
Under this assumption, it can be shown that $Y$ is distributed as a Gumbel variable
\beq\label{eq:gumbel}
P_Y(y) = e^{y - e^y}\, .
\eeq
The Gaussianity can be viewed as a consequence of the central limit theorem applied to the sum of levels \cite{cipolloni2023functionalcentrallimittheorems,Cipolloni_2023}.
In fact $Z$ is the sum of an exponential number of complex variables (exponential in the number of degrees of freedom of the physical system). In any situation in which the temperature is sufficiently high to ensure that a large enough number of terms contributes to the sum and such that one may assume that these values are independent, or more generally that two successive partial sums $\sum_1^k e^{-(\beta+it)e_i}$ and  $\sum_{k+1}^{2k} e^{-(\beta+it)e_i}$ are independent for some $k$, then the central limit theorem applies. A recent paper discusses this result in a more general context as the outcome of a random walk in the complex plane \cite{venuti2025integrabilitychaosfractalanalysis}.

At this point we can address the question of what are the fluctuations of $Y$ which amounts to study the fluctuations of the quenched $\mathrm{SFF}$, knowing that $\tilde{Z}(\beta+i t)$ is Gaussian. As we explain in the next section it turns out that the large negative fluctuations are the close-by zeros of the partition function considered in \cite{bunin2024fisher}.

Note that the statistics of the variable $Y$ which follows for the assumption of Gaussianity of $Z$ is independent on the chaoticity of the model, which is instead contained in the shape of $\langle \mathrm{SFF}(t)\rangle$ (having or not a ramp).
In the following in fact we will check this distribution in several models: (i) a system with Poissonian energy levels, (ii) a many-body spin glass in the chaotic high temperature phase and (iii) a non-hermitian version of the same many body problem.

\section{Large fluctuations of the quenched SFF and Fisher zeroes}\label{Sec_fluct}

The general form of the fluctuations of the function $Y(\beta+i t)$ may be understood well in the complex plane. As already mentioned the {\it ramp} and  the {\it plateau} regimes occur when 
the fluctuations of $Z(\beta+i t)$ are larger that its average and in  this regime of large times one encounters a region with a distribution
of zeroes of $Z(\beta+it)$.
On the other hand, Fisher zeros close to the real axes have been studied in several works, see for instance, Refs~\onlinecite{falcioni1982complex,marinari1984complex,obuchi2012partition}.

Large negative  fluctuations, the exponential tail of the Gumbel distribution, may be 
understood in a simple way as `near misses' of the $\beta+i\,t$ line of zeroes of $Z$. 
In fact if one plots $\log \mathrm{SFF}(t)$ versus $t$ for one instance, focusing for instance on the plateau they will find spikes which have been interpreted as the proximity to a nearby zeros of the partition function in complex temperature \cite{bunin2024fisher}. 
In Section \ref{SecVI} we will show the occurrence of one of these big fluctuations in a hermitian many-body system.
Let us assume that we have $M$ zeros uniformly distributed in a region of phase diagram $\Delta\beta$ and $\Delta t$. We suppose to be in the plateau and to move along a line of length $\Delta t$. Close to one of these zeros the partition function can be expanded up to linear order so that, {\it in a spike},  the variable $Y$ approximately reads $Y = \log( x^2+z^2)$ with $x=\beta-\beta_i$ and $z=t-t_i$
where $\beta_i$ and $t_i$ are the coordinates of the $i$-th nearby zero. By assumptions, the zeros are uniformly distributed
and if we sample in time randomly, just requiring to be close enough to a zero, both variables $x$ and $y$ can be thought to be uniformly distributed within some range. This automatically implies that the variable $Y$ is exponentially distributed $P_{Y}(y) = e^{y}$ for $y \ll 0$, exactly as in the tail of the Gumbel distribution.
An alternative study could be to
collect statistics of all local minima (in time) $Y_m$ of $Y = \log \mathrm{SFF}_{\beta}(t) - \log {\mathcal D}$, which, for small enough values of $Y_m$, occur when we go close to a zero representing a spike. These spikes occur when the time we consider coincides with the time of one of the nearby zeros, so that $t=t_i$. The value of $Y_m$ will be dominated by the distance between the zero and the line along the time direction that we follow, which is $\log x^2 = \log|\beta-\beta_i|^2$. Using the same assumptions, the variable $Y_m$ will be exponentially distributed and, in particular
\beq\label{PQ}
P_{Y_m}(y_m) \propto e^{y_m/2} \qquad\text{for $y_m \ll 0$} .
\eeq
In Section \ref{SecVI} we will provide an accurate verification of both these results. This argument highlights that the exponential tail in the distribution of $Y$ given in Eq.~\eqref{eq:gumbel} is tightly related to the presence of spikes in the time sequence, but this discussion of exponential tails in the distribution of local minima determined by the presence of nearby zeros seems to us complementary to the Gaussianity approach with which the Gumbel distribution has been obtained.

 We will show that similarly to the hermitian case one has large downward fluctuations also in the non-hermitian case. In this context, the DSFF can be viewed at finite system size as an analytic function of two complex variables $\beta_1+i t_1$ and $\beta_2+i t_2$. It is known that the zeros of multivariate analytical complex functions lie on hypersurfaces and are not isolated. In our definition of DSFF we have chosen to fix $\beta_1=\beta_2=0$ and upon changing $t_1$ and/or $t_2$ one can encounter such manifolds and touch a zero. In the plane $t_1-t_2$ the zeros are isolated as we see by plotting the variable $Y$ in a range of time in the plateau. Close to one of the zeros one can expand $Z$ in $t_1$ and $t_2$ up to linear order and similar arguments to the single complex variable in the hermitian case would follow.

In Section \ref{SecVI} we show the occurrence of one of these big fluctuations in a non-hermitian many-body system.




\section{Results for many-body systems}\label{SecVI}

\subsection{Poissonian spectrum: the random energy model}

We start by considering the simplest case of ``Poissonian spectra", meaning i.i.d. energy levels. 
If one considers the consecutive gap ratios:
\begin{equation}
    \label{eq:gapratiosHerm}
    r_n=\frac{E_{n+1}-E_n}{E_n-E_{n-1}}.
\end{equation}
there is an analytical prediction for uncorrelated energies which reads
\begin{equation}
    \label{eq:consgapPoisson}
    P(r)=\frac{1}{(1+r)^2}
\end{equation}
which our model satisfies.

\begin{figure}
    \centering
    {\includegraphics[width=8.25cm]{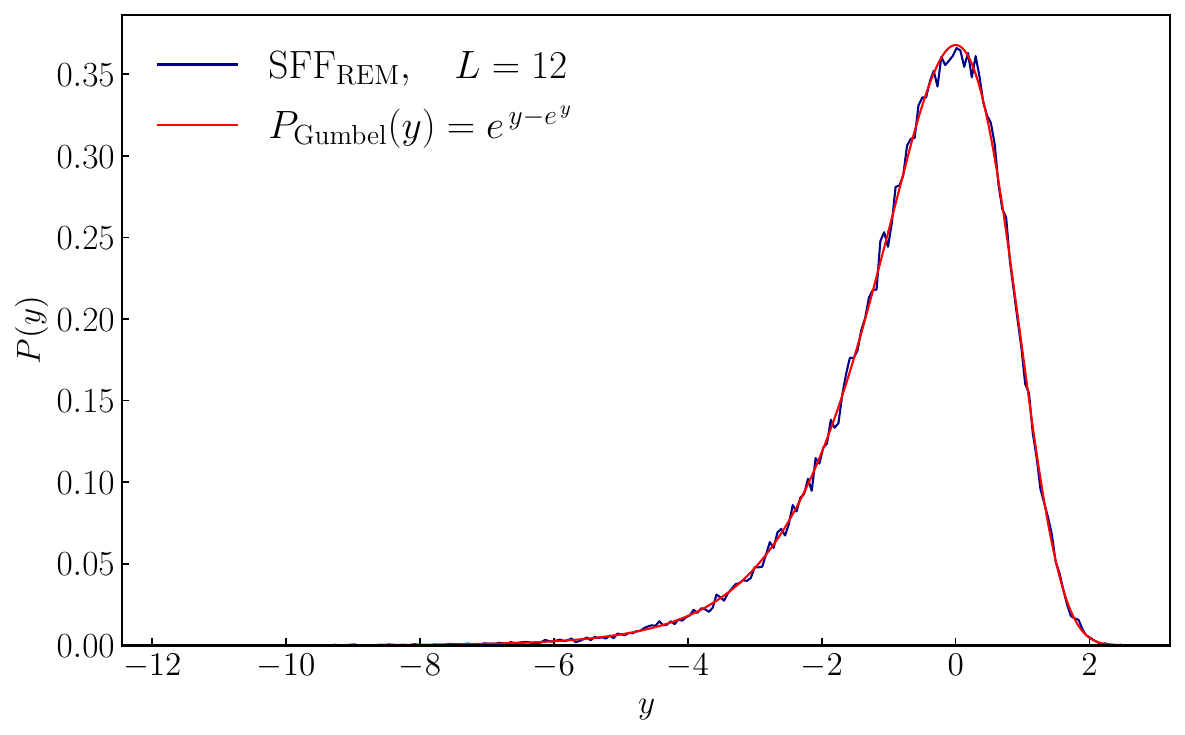}}
    \caption{Empirical Distribution of the variable $Y$ where the statistics are taken for a specific time over $10^5$ samples of spectra of the REM with $L=12$. The line in red corresponds to the Distribution of the variable $Y$ for $Z(\beta+it)\sim\mathcal{N}(0,\sigma^2)$.}
    \label{fig:Poissonian Spectrum - Gumbel Instance Averaging (Plateau)}
\end{figure}

\begin{figure}
    \centering
    {\includegraphics[width=8.4cm]{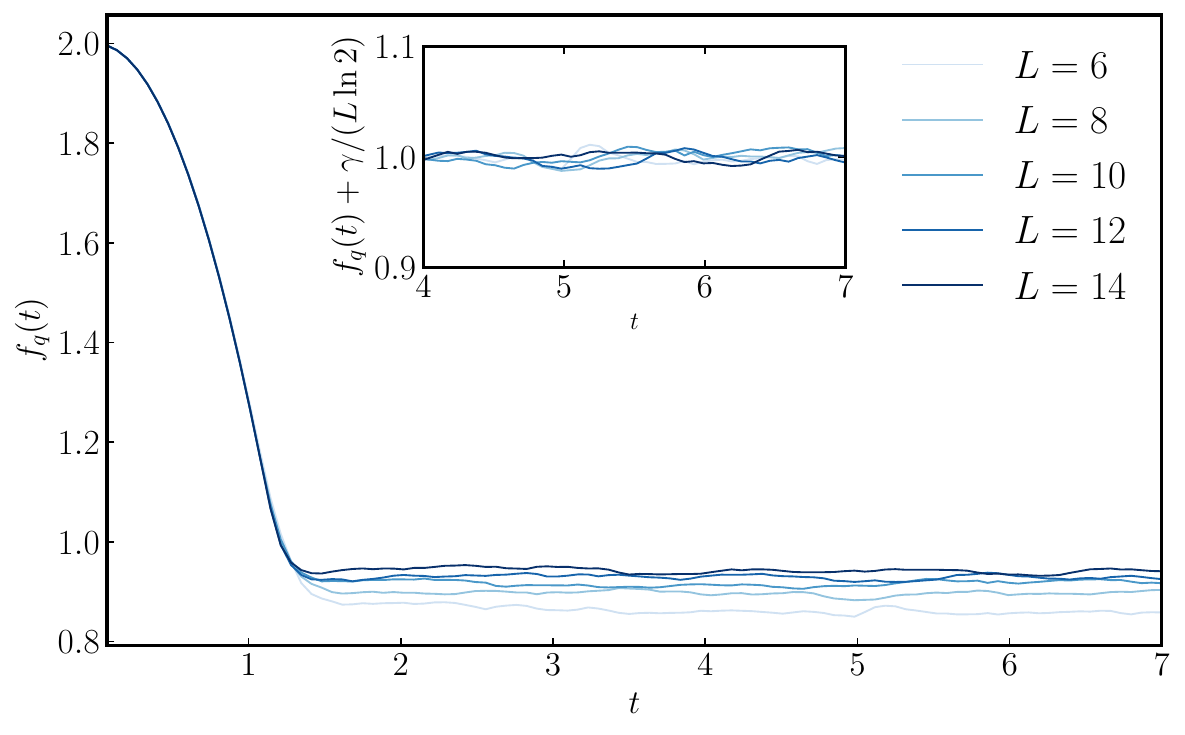}}
    \caption{{\color{black}Time sequence of the quenched average (1000 samples) of the $\mathrm{SFF}(t)$ for the REM using various system sizes and $\beta=0$. Adding the quenched/annealed correction of Eq.~(\ref{eq:QA}) in the inset, we see that the different curves match at the plateau given in Eq.~(\ref{SFF_quenched_REM}).}
    }
    \label{fig:SFF_REM}
\end{figure}

An archetypal such case, which displays a rich phase diagram in (complex) temperature, is the Random Energy Model (REM) \cite{derrida1981random}, which can be defined through its spectrum; $2^L=\mathcal{D}$ i.i.d. energy levels drawn at random from a Gaussian law {\color{black} $\sim \mathcal{N}(0,L/2)$}. 
For this model we tested the validity of the Gaussian assumption \cite{kabluchko2014complex} by obtaining the empirical distribution of $Y$ by sampling the $\mathrm{SFF}$ which we calculated at one time in the plateau over different disorder realizations. The result is shown in Fig.~\ref{fig:Poissonian Spectrum - Gumbel Instance Averaging (Plateau)}
It is well known that the ensemble average of the SFF obtained over such spectra should exhibit an early-time slope which depends on the details of the system (the distribution of the levels in this case), followed by a plateau. For the REM the averaged SFF at $\beta=0$ can be shown to be
\begin{equation}
    \label{eq:REMSFF}
    \begin{gathered}
      \langle \mathrm{SFF}(t) \rangle=\mathcal{D}+\mathcal{D}(\mathcal{D}-1)e^{-L\frac{t^2}{2}},\\
       \kappa_c(t)=\mathcal{D}(1-e^{-L\frac{t^2}{2}}).
    \end{gathered}
\end{equation}
The connected correlations starts from zero as it is always the case but with an increasingly rapid rate as the size increases, it saturates to a constant, signaling the absence of correlations between energy levels.
The quenched SFF in the thermodynamic limit has been calculated by Derrida \cite{derrida1991zeroes} which for $\beta \leq \beta_c/2 = \sqrt{\ln 2}$ reads:
\beq\label{SFF_quenched_REM}
\lim_{L\to\infty} f_q(t) = {\color{black}\frac{1}{\ln 2}
\begin{cases}
2 \ln 2 + \frac{1}{2} (\beta^2-t^2) \qquad \text{for $t<t_c$} \\
\ln 2 + \beta^2  \qquad\qquad\qquad \text{for $t>t_c$}
\end{cases}}
\eeq
and $t_c=\sqrt{2 \ln 2-\beta^2}$.
In Fig.~\ref{fig:SFF_REM} we show this result and in the inset the fact that the plateau has corrections $-\frac{\gamma}{(L\ln 2)}$ as in Eq.~(\ref{eq:QA}).
Eqs.~(\ref{eq:REMSFF}) and (\ref{SFF_quenched_REM}) show that at high enough temperature in the complex plane, quenched equals annealed as we anticipated, despite the large fluctuations of the SFF.

\subsection{Hermitian matrices}

We now move to consider a ``chaotic" XY spin glass at high temperature defined as
\begin{equation}
    \label{eq:XYSPINGLASS}
    \begin{array}{ll}
    \displaystyle
    H &   \displaystyle= \sum_{i<j}J_{ij}\left(S_i^xS_j^x+S_i^yS_j^y\right) 
    \\ \vspace{-0.2cm} \\
    \displaystyle 
   &   \displaystyle=  \sum_{i<j}\frac{J_{ij}}{2}(S_i^+S_j^-+ S_i^-S_j^+) \, 
    \end{array}
\end{equation}
where the coupling constants $\{J_{ij}\}$ are independent identically distributed random variables $\sim \mathcal{N}(0,1/\sqrt{L})$ and $\{S_i^{\mu}\}_{\mu}$ are the Pauli matrices. The Hamiltonian commutes with the magnetization operator $m=\frac{1}{L}\sum_{i}S_i^z$. As such, we constrain ourselves to one symmetry sector of the entire Hilbert space $\mathcal{H}$, i.e., a subset of states with the same value of a conserved quantity, in this case the magnetization, which we take to be $m=2/L$ unless stated otherwise. The symmetries of the system, namely time reversal invariance, imply that it will follow GOE statistics. This  model was examined in Ref.~\onlinecite{bouverot2024random} under the prism of the Eigenstate Thermalisation Hypothesis \cite{srednicki1999approach}.

\begin{figure}
    \centering
    {\includegraphics[width=8.4cm]{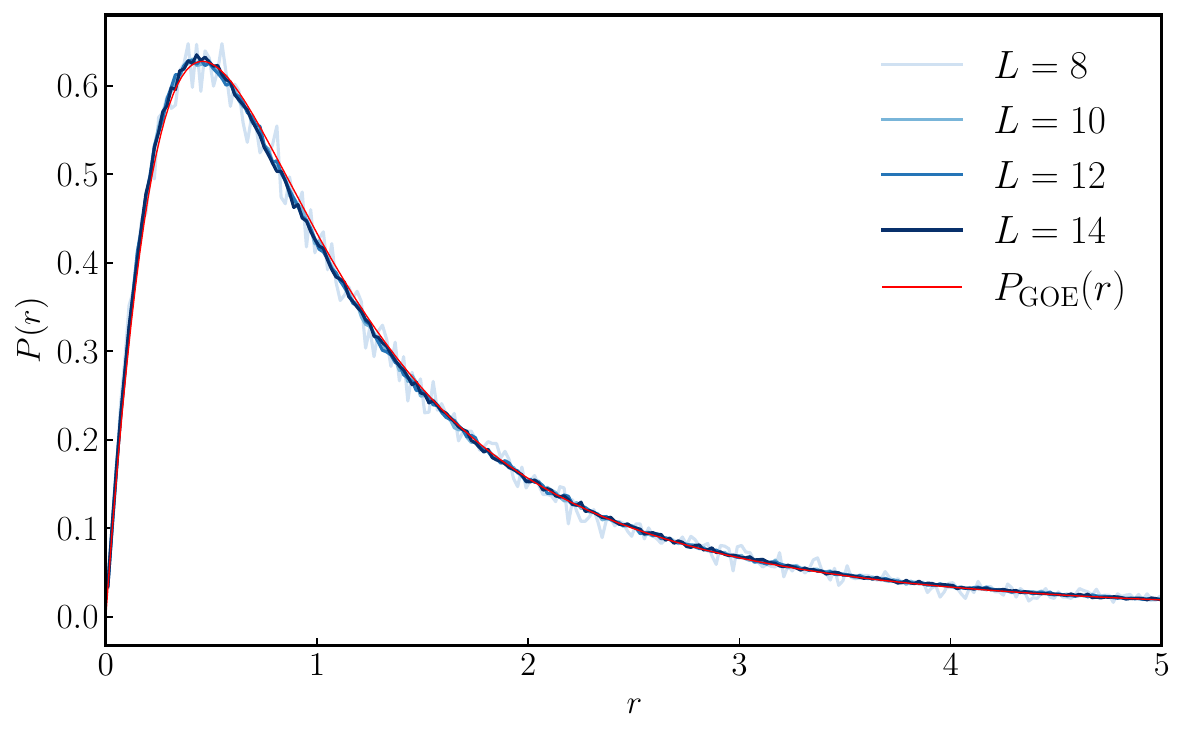}}
    \caption{Empirical Distributions of the spectral gap ratios in the bulk of the spectrum of Eq.~(\ref{eq:XYSPINGLASS}) for different system sizes and symmetry sectors using $10^5$ eigenvalues. In order of increasing system size, we use: $L= 8,10,12,$ and $14$.
    }
    \label{fig:Gapratiosmany}
\end{figure}

A standard way to argue about the chaoticity of the model is by studying consecutive gap ratios~\cite{PhysRevLett.110.084101} given in Eq.~(\ref{eq:gapratiosHerm}).
Considering the ratio of energy gaps instead of the gaps themselves presents the advantage of bypassing the process of unfolding the spectrum, as the $\{r_n\}$ are independent of any fluctuations in the spectral density that are system (and disorder instance) dependent. In the case of the GOE, the analytical expression (up to small deviations) for the distribution of the $\{r_n\}$ is given by:
\begin{equation}
    \label{eq:gapratios formula}
    P(r)=\frac{27}{8}\frac{(r+r^2)}{(1+r+r^2)^{5/2}}.
\end{equation}
In Fig.~\ref{fig:Gapratiosmany} we show the accuracy of this result for the model in Eq.~(\ref{eq:XYSPINGLASS}).

\begin{figure}
    \centering
    {\includegraphics[width=8.25cm]{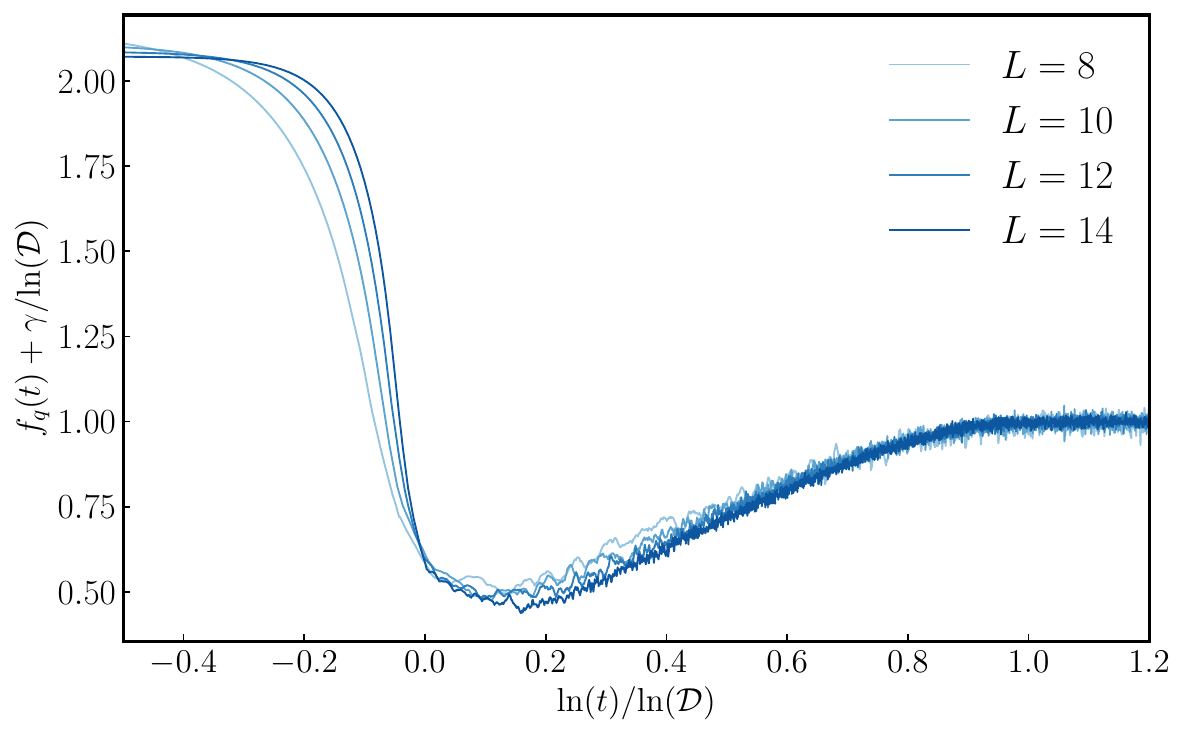}}
    \caption{Time sequence of the quenched average of the $\mathrm{SFF}$ for different system sizes in Eq. (\ref{eq:XYSPINGLASS}). In order of increasing $L$, we use: $L= 8,10,12,$ and $14$. To have the curves match, we plot the normalized quenched average against $\ln{(t)}/\ln{(\mathcal{D})}$ and include the sub-dominant correction in Eq. (\ref{eq:QA}).}
    \label{fig:SFF_XY}
\end{figure}

As we discussed a time resolved method to look at the chaoticity of the system is the spectral form factor.
Both the annealed and the quenched spectral form factor for high enough temperature exhibit a clear linear ramp, accompanied by a plateau at late times. Fig.~\ref{fig:SFF_XY} shows that this behavior is consistent for different system's sizes.
At low temperatures the model is expected to undergo a spin-glass transition.
In this regime annealed and quenched averaged are distinct and one should study the quenched one as annealed averaged are dominated by rare realisations of the disorder. 

\begin{figure}
    \centering
    {\includegraphics[width=8.4cm]{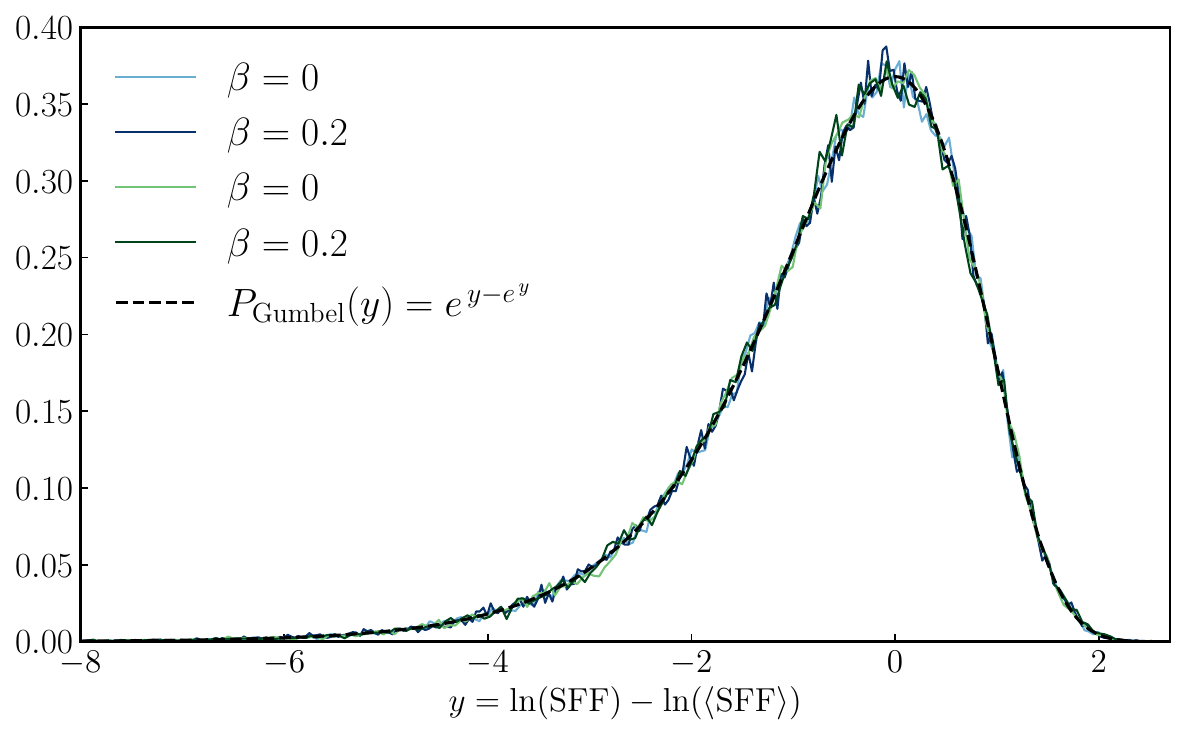}}
    \caption{Empirical distribution of the $Y$ variable for various cases. The blue lines are obtained using samples of the $\mathrm{SFF}(t)$ calculated at $10^5$ times in the plateau with $L=14$. In order of increasing shades of blue, we use $\beta=0,0.2$. The green shades are obtained by sampling $5\times10^4$ disorder instances using a single time in the plateau of the $\mathrm{SFF}$ and setting $L=10$.
    }
    \label{fig:GUMBELXY}
\end{figure}

Next, for this model, we study the distribution of the variable $Y$ and confirm that it is indeed distributed as Eq.~(\ref{eq:gumbel}) when we sample either over times in the plateau of the $\mathrm{SFF}$ or over disorder realizations. The existence of a spin glass transition implies that we expect this behavior to break down for large enough $\beta$. Some findings are shown in Fig.~\ref{fig:GUMBELXY}. 

\begin{figure}
    \centering
    {\includegraphics[width=8.5cm, height= 5.27215189873cm]{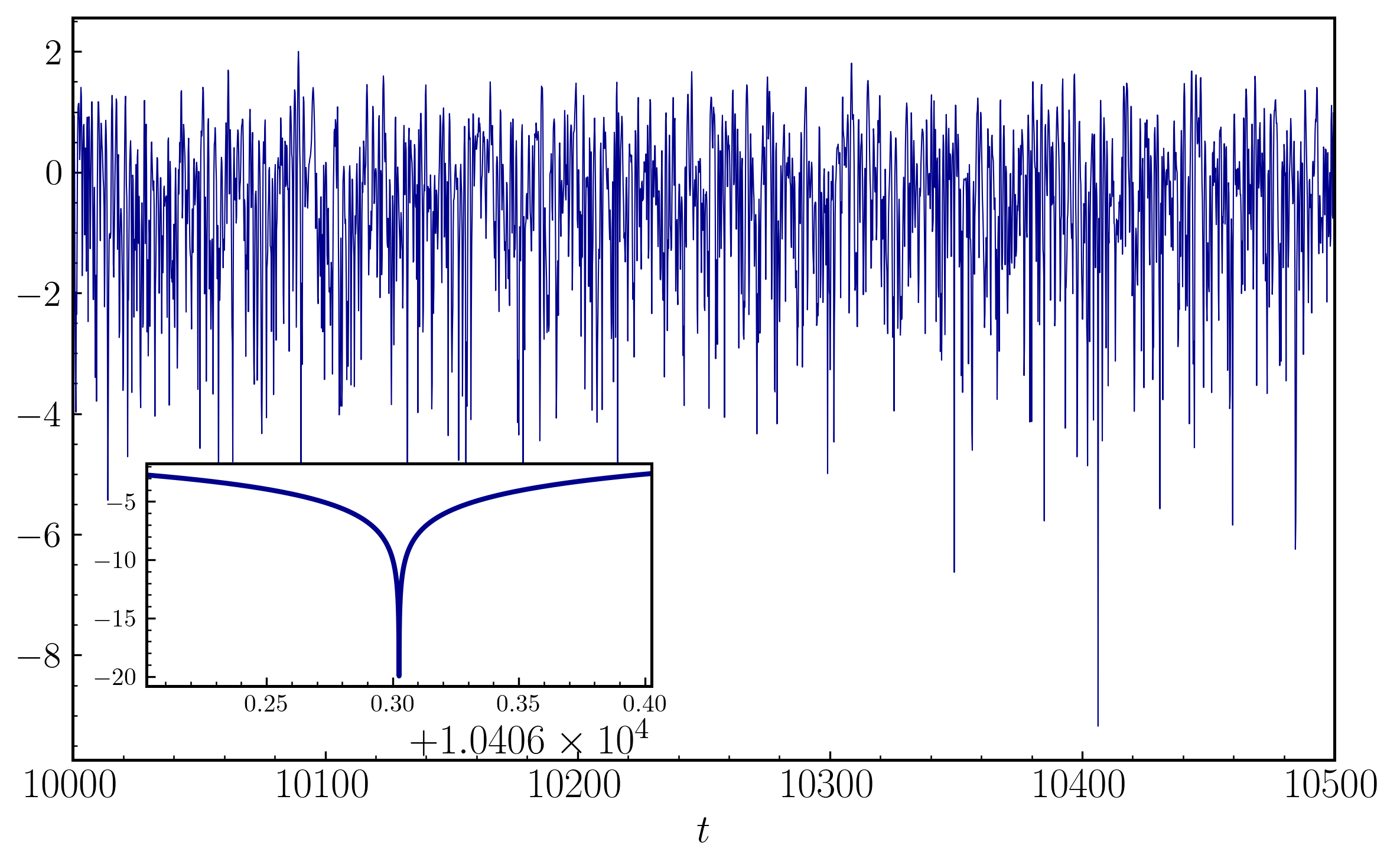}}
    \caption{Time sequence of the variable $Y$ for the  XY model in Eq. (\ref{eq:XYSPINGLASS}) with $L=14$.
    The $\mathrm{XY}$ (with $\beta=0.2$ here) clearly shows a ramp, but when one considers the $Y$ variable, the ramp is not visible and the properties of $Y$ are the same as in the plateau.
    After spotting the deepest spike (which also corresponds to a spike in the $\mathrm{SFF}$ itself), we ``zoom" around it in the inset, and by slightly varying $\beta$, we can make the spike arbitrarily deep.
    }
    \label{fig:LNSFF}
\end{figure}

Fig.~\ref{fig:LNSFF} shows the evolution of the variable $Y$ at late times and in the inset we show one deep spike that is found by slightly moving in the complex plane. In fact the possibility of going so deep in value is allowed by tuning the temperature and therefore approaching the zero of the partition function.

\begin{figure}
    \centering
    {\includegraphics[width=8.7cm]{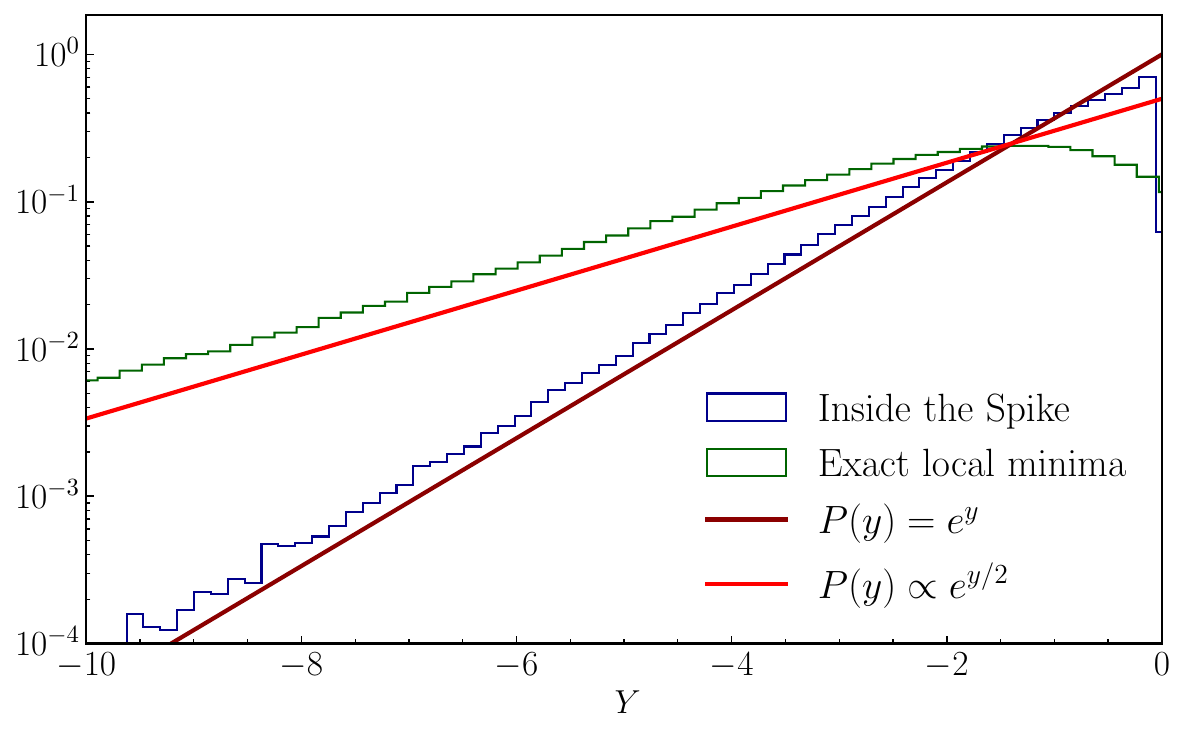}}
    \caption{Empirical distribution for the local minima of the $\mathrm{SFF}$ for system size $L=8$ shown in green. The samples are obtained from a large time interval in the plateau using $\Delta t=5\times10^{-3}$. The blue line corresponds instead to samples taken from a random time inside each spike, rather than the exact local minimum.  
    }
    \label{fig:Local}
\end{figure}

By calculating one instance of the $\mathrm{SFF}(t)$ for a long enough time window in the plateau, we can numerically verify the assertions of Sec.~\ref{Sec_fluct}. One should discretise the time well enough to be sure to locate precisely the local minima (local in time). After tracking all the local minima we obtain their empirical distribution which is compatible with Eq.~(\ref{PQ}).
We then implement a second procedure which allows us to draw at random a time ``inside a spike" in order to select a large fluctuation but not the precise minimum. In order to do this we define a range around each local minimum, which we consider to be the size of the spike. Selecting one random point within this range amounts to relaxing the condition to a near miss in both time and temperature, revealing the tail of the Gumbel distribution. 
These two results are shown in Fig.~\ref{fig:Local}.

\subsection{Non-hermitian matrices}\label{subsec:Non-Herm-Many-Body}

As a model of a dissipative system, we consider a non-Hermitian generalization of the Hamiltonian defined in Eq.~(\ref{eq:XYSPINGLASS})
\begin{equation}
    \label{eq:DISSXYSPINGLASS}
    H = \sum_{i<j}\frac{J_{ij}}{2}(e^{-g}S_i^+S_j^-+e^{g}S_i^-S_j^+),
\end{equation}
where $g$ parametrizes the strength of non-hermiticity. Since the Hamiltonian is still real, time reversal symmetry persists, and now we conjecture that the system will behave like a GinOE matrix. 

\begin{figure}
    \centering
{\includegraphics[width=8.25cm]{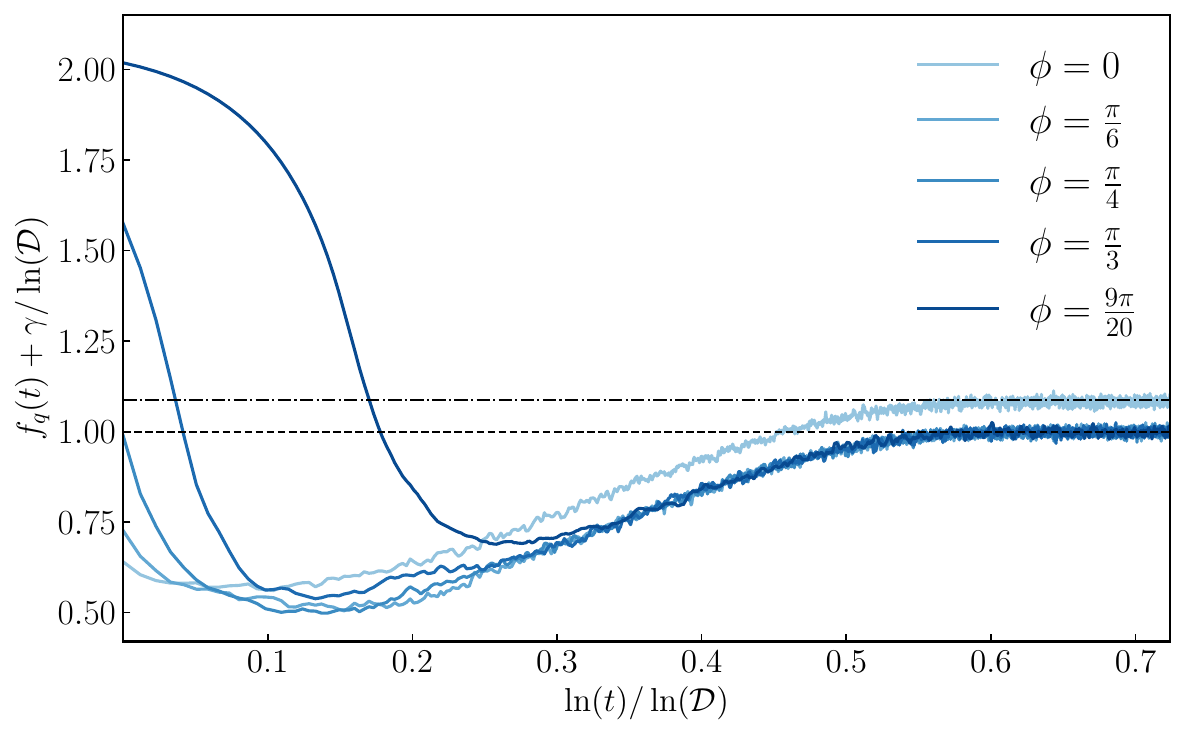}}
    \caption{Time sequence of the quenched average of the $\mathrm{DSFF}$ for 400 disorder instances of the Hamiltonian in Eq. (\ref{eq:DISSXYSPINGLASS}) with $L=14, m =2/L, g=0.4$ for various angles $\phi$. Including the correction in Eq. (\ref{eq:QA}) and normalizing by $\ln\mathcal{D}$ means that the plateau is expected to be at height 1 (black dashed line). For $\phi=0$, the $\mathrm{DSFF}(t)$ contains the doubly degenerate complex conjugate eigenvalue pairs, implying that the plateau value should be $\ln(2\mathcal{D)}/\ln\mathcal{D}$ (dotted dashed line). }\label{fig:DSFF_XY_3003_g_0.4}
\end{figure}

In Fig.~\ref{fig:DSFF_XY_3003_g_0.4} we show the quenched average for the DSFF of model given in Eq.~\eqref{eq:DISSXYSPINGLASS} and we observe a polynomial ramp. The inherent lack of rotational isotropy of the spectrum of the non-Hermitan Hamiltonain [Eq.~\eqref{eq:DISSXYSPINGLASS}] in the complex plane 
makes a quantitative comparision between the numerically obtained $\mathrm{DSFF}$ and the analytical prediction from RMT ill-suited due to angular dependence in the early time regime.

\begin{figure}
    \centering
    {\includegraphics[width=8.25cm]{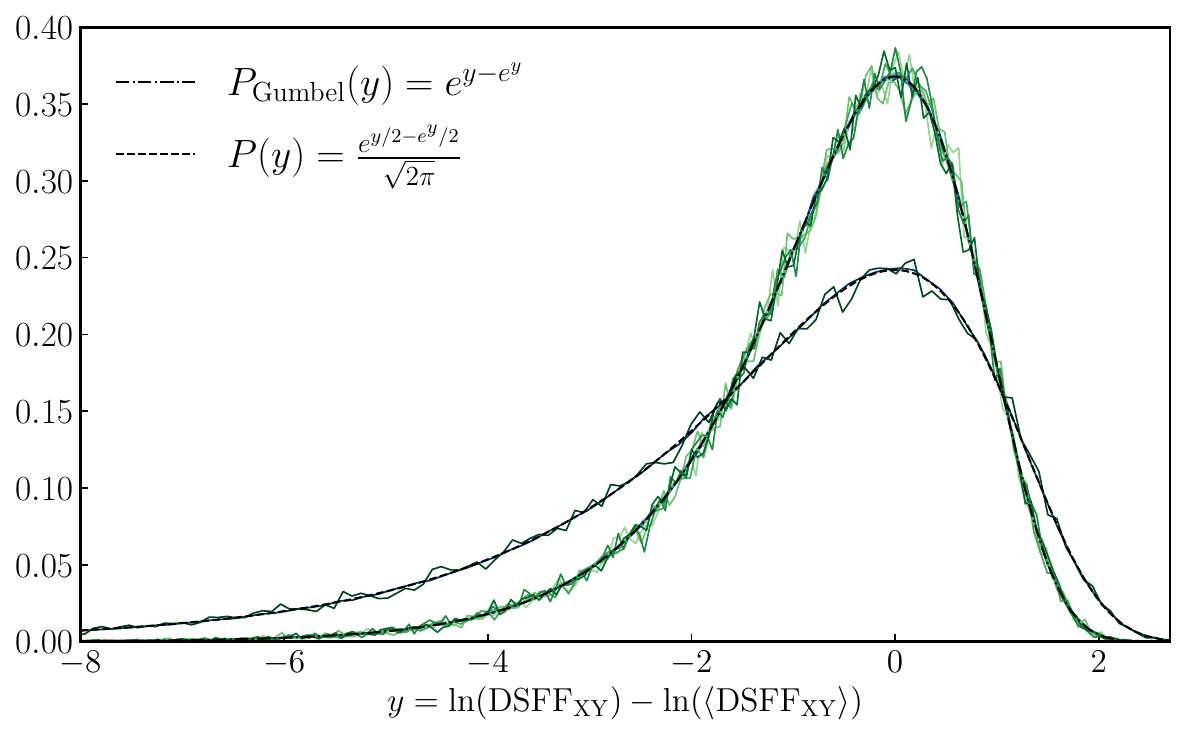}}
    \caption{Distribution of the variable $Y$ for a multitude of cases. We calculate the $\mathrm{DSFF}$ using $L=14,m=2/L,g=0.4$ for $\phi=0,{\pi}/{6},{\pi}/{4},{\pi}/{3},{9\pi}/{20},$ and ${\pi}/{2}$. Blue lines are obtained by calculating the $\mathrm{DSFF}(t)$ at $\approx 2.5\times10^6$ time points in the plateau, having set $\Delta t=1$. Increasing shades of each color always indicate an increasing angle. In all cases when $\phi=\frac{\pi}{2}$, we remove the real eigenvalues before any calculation. Then, for some specific time in the plateau (green), we sample over $5\times10^4$ disorder instances, hence the increased noise. For these averages, we set $L=10$. Averaging over disorder samples for a time in the ramp works in all cases here except $\phi=\frac{\pi}{2}$, where the ramp is not well defined (for this system size). 
    }
    \label{fig:GumbelDissipativeXY}
\end{figure}

Also here, the  (complex) Gaussianity assumption is obeyed by the DSFF, or better by $Z(t_1,t_2)$, whether one samples over times in the plateau or over disorder instances for a single late time. This holds for all angles except for  $\phi=\frac{\pi}{2}$, as it is shown in Fig~\ref{fig:Gumbel_ALL} and Fig.~\ref{fig:GumbelDissipativeXY} for the non-Hermitian RMT ensembles and the Hamiltonian in Eq.~(\ref{eq:DISSXYSPINGLASS}) respectively. Projecting on the imaginary axis suggests a modification of the Gaussianity assumption; it now holds that $Z(t_1,t_2)=C+C^*+A$, where $C\sim\mathcal{N}(0,\sigma^2)$ 
is complex, and $A$ is equal to the number of real eigenvalues (itself a random variable for the GinOE). Removing the real eigenvalues for the sake of simplicity leads to the distribution for $Y$ being $P(y)=e^{y/2-e^y/2}/\sqrt{2\pi}$. In Appendix \ref{App_RMT} we discuss the validity of this analysis in the context of non hermitian random matrices, while here we focus on non-Hermitian quantum many-body systems.

\begin{figure}
    \centering
    {\includegraphics[width=8.5cm, height= 5.27215189873cm]{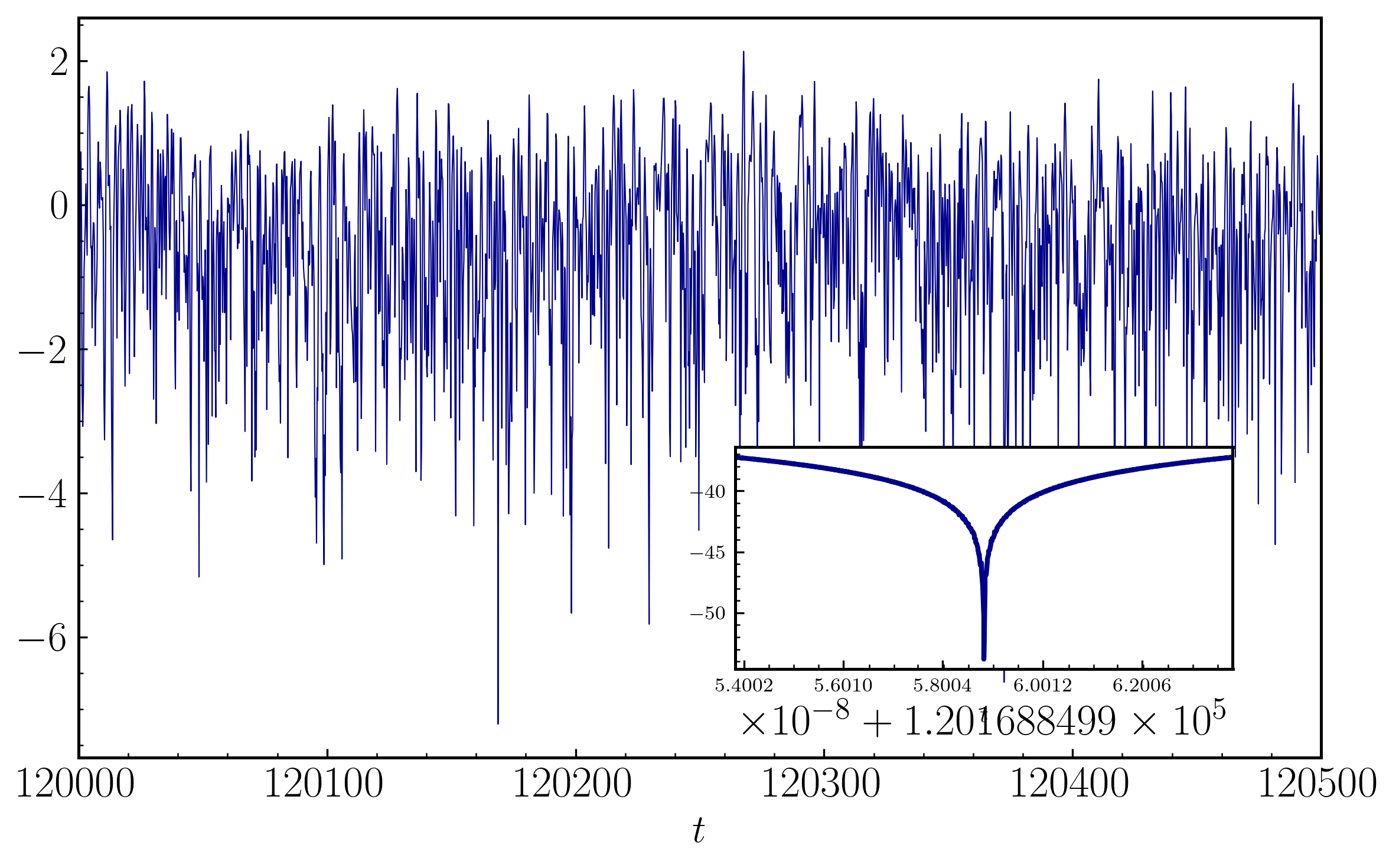}}
    \caption{Time sequence of the variable $Y$ for the non hermitian XY model in Eq. (\ref{eq:DISSXYSPINGLASS}) using $L=14,m=2/L,$ and $g=0.4$. We start from a single instance of the DSFF with $\phi=\pi/4$ deep in the plateau. After finding the deepest spike, we repeatedly recalculate the DSFF around it by varying $\phi$, keeping the value for which the DSFF is minimized, and repeating by decreasing the relative variation in angle. Proceeding up to the point allowed by default machine precision ($\Delta\phi/\phi\approx10^{-16}$) allows us to obtain a minimum of $Y\approx-55$, $\text{DSFF}(t)\approx 10^{-22}$.
    }
    \label{fig:DSFF_spike}
\end{figure}

Similar to the hermitian case the log of the SFF or equivalently the variable $Y$ has deep spikes downward.
In Fig.~\ref{fig:DSFF_spike} we show the time dependence of one instance of $Y$ and in the inset we show how upon tuning $\phi$ slightly, it is possible to go close to a zero.
Note that we can view the DSFF as a complex function of two complex variables, where in this case we have set $\beta_1=\beta_2=0$. The zeros of such analytic function are more complex than those of a single variable but as shown in Fig.~\ref{fig:DSFF_spike} we find that they are similar to the hermitian case.

\begin{figure}
    \centering
    {\includegraphics[width=8.25cm]{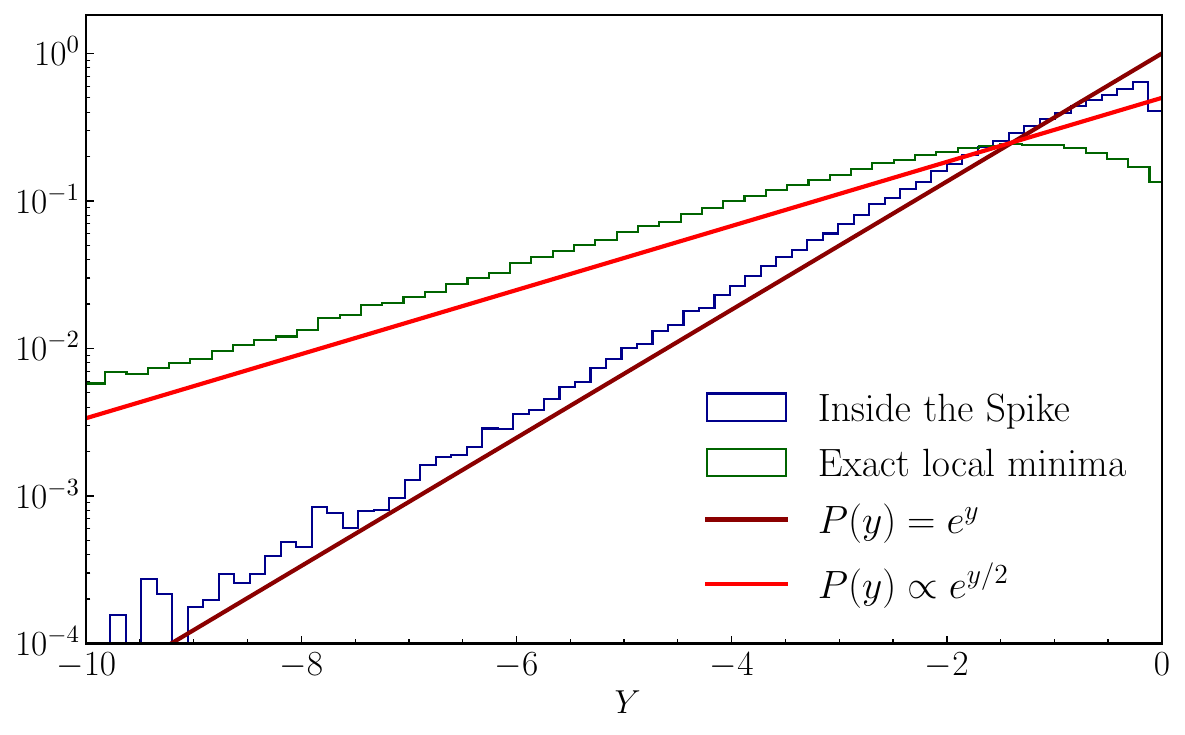}}
    \caption{Empirical distribution for the local minima of the $\mathrm{DSFF}$ for system size $L=10$, $m=2/L,$ and $g=0.4$ shown in green. The samples are obtained from a large time interval in the plateau using $\Delta t=5\times10^{-3}$. The blue line corresponds instead to samples taken from a random time inside each spike, rather than the exact local minimum. We set $\phi=\pi/4.$}
    \label{fig:DISS_LOCAL}
\end{figure}

In Fig~\ref{fig:DISS_LOCAL}, similar to Fig~\ref{fig:Local}, we provide further evidence towards the interpretation of the spikes as near-miss of zeros in the non-hermitian case
and the large negative fluctuations of the variable $Y$ as one of them. The figure shows the  distributions of the true local minima and that of values at random times inside each spike and the two agree with the discussion in Sec~\ref{Sec_fluct} and are the same as in Fig.~\ref{fig:Local}.

\begin{figure}
    \centering
    {\includegraphics[width=8.25cm]{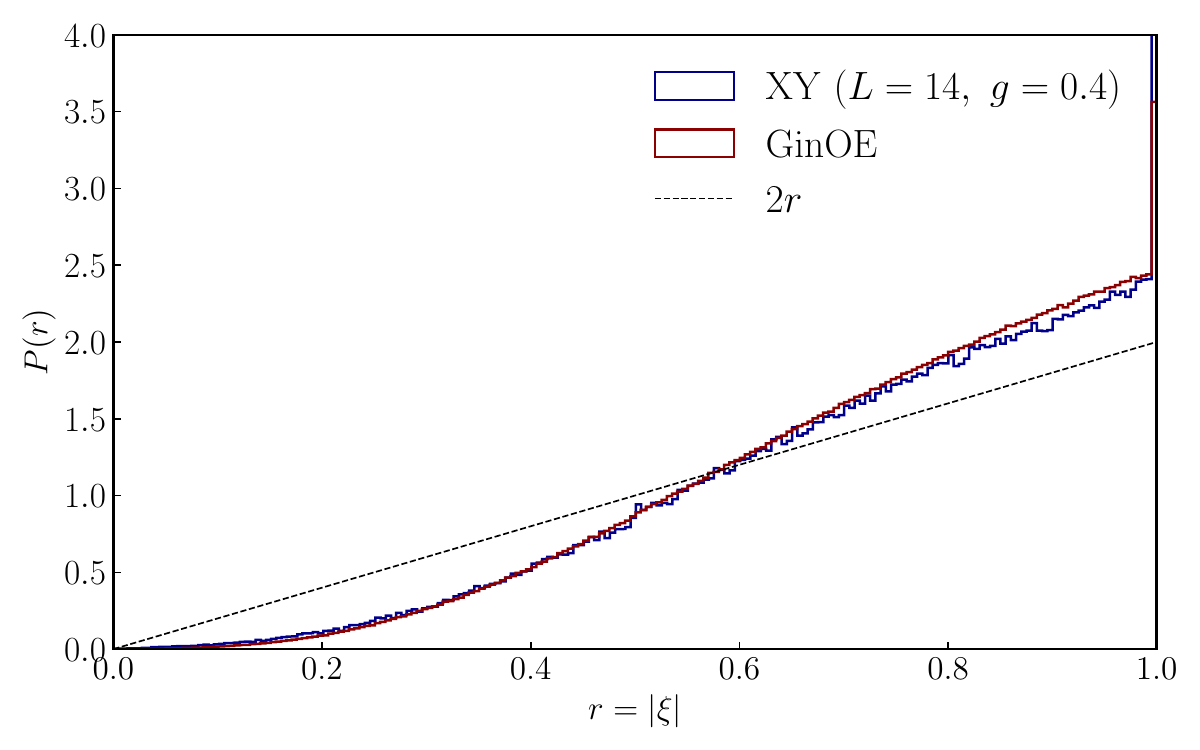}}
    \caption{Empirical marginal distribution of $r_n$ in Eq. (\ref{eq:complexpsacingratio}) obtained from 315 samples of the Disspative XY model in Eq. (\ref{eq:DISSXYSPINGLASS}). The black dashed line corresponds to the distribution for uncorrelated levels.}
    \label{fig:TRIPLE COMPARISON RADIAL DISSIPATIVE XY}
\end{figure}

\begin{figure}
    \centering
    {\includegraphics[width=8.25cm]{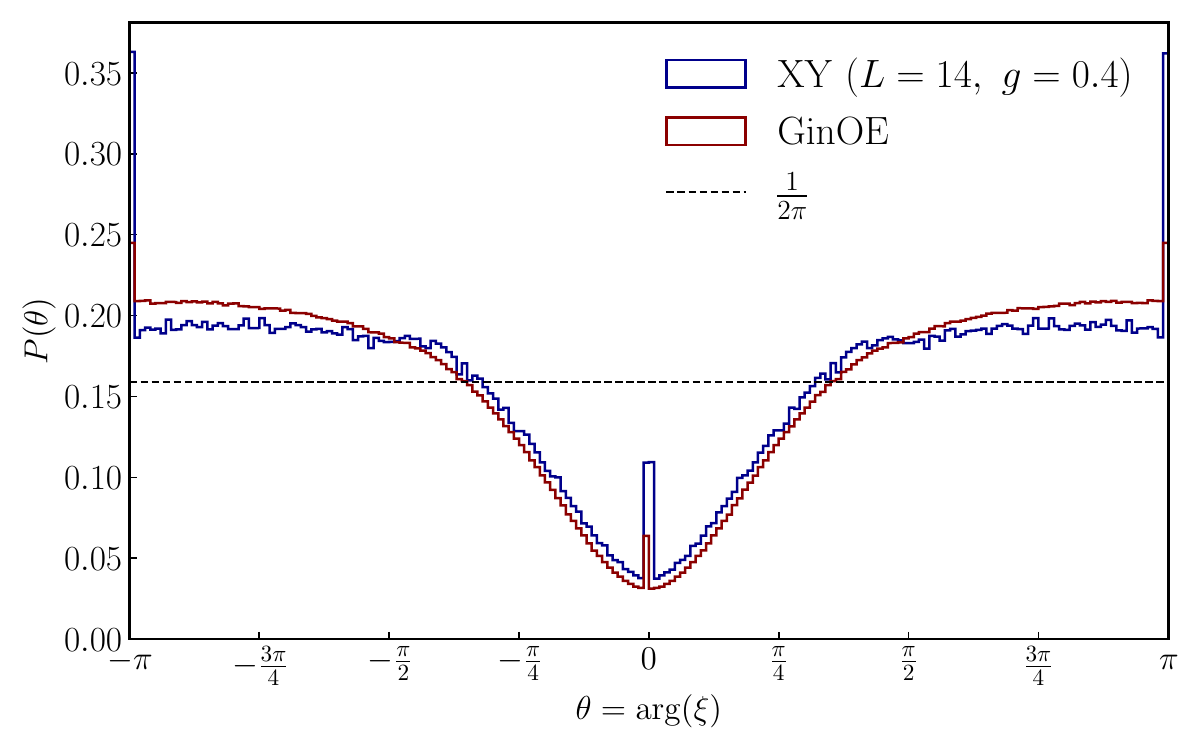}}
    \caption{Empirical marginal distribution of $\theta_n$ in Eq. (\ref{eq:complexpsacingratio}) obtained from 315 samples of the Disspative XY model in Eq. (\ref{eq:DISSXYSPINGLASS}). The distribution in the case of uncorrelated levels is uniform over $[-\pi,\pi]$.}
    \label{fig:TRIPLE COMPARISON ANGULAR DISSIPATIVE XY}
\end{figure}

We complete the discussion by analyzing the spacing between eigenvalues. One possible extension of Eq.  (\ref{eq:gapratios formula}) for non-Hermitian eigenvalues is the complex spacing ratio that was introduced in Ref.~\onlinecite{PhysRevX.10.021019}:
\begin{equation}
    \label{eq:complexpsacingratio}
    \xi_n=\frac{\lambda_n^{NN}-\lambda_n}{\lambda_n^{NNN}-\lambda_n}=r_ne^{i\theta_n},
\end{equation}
where $\lambda_n^{NN},\lambda_n^{NNN}$ are the nearest and next-to-nearest neighbors of the eigenvalue $\lambda_n$ in terms of the absolute distance measured on the complex plane. For comparison's sake, we focus on the radial and angular marginal distributions of $\xi$ presented in Fig~\ref{fig:TRIPLE COMPARISON RADIAL DISSIPATIVE XY} and Fig~\ref{fig:TRIPLE COMPARISON ANGULAR DISSIPATIVE XY} repectively. The marginal distributions for the GinOE were numerically obtained in this case for an ensemble of matrices with $\mathcal{D}\approx3500$.

\section{Conclusions}\label{SecVII}

We contrast the definition of the annealed and the quenched SFF. While the SFF is notoriously not self averaging the quenched SFF displays bounded fluctuations.

In a time sequence one instance of the logarithm of the SFF displays big and ``thin" downward spikes associated with the proximity of a zero of the partition function. The distribution of these spikes is exponential and we give a simple argument for this result. Such distribution encodes the large negative fluctuations of the quenched SFF in the ramp-plateau regime, where it is known to be the modulus squared of a complex Gaussian random variable. 
In fact, the exponential behaviour is compatible with the tail of the Gumbel distribution which is expected by studying the statistics of the logarithm of the SFF in the ramp-plateau regime and compares remarkably well with our data for many-body systems.

We extend this discussion to the non hermitian case discussing a non-hermitian generalisation of a chaotic spin glass.
In particular also in this case we find that a single sample of the logarithm of the SFF is characterized by big downward spikes whose statistics is also exponential. In addition to non-Hermitian hamiltonians, we also discuss non-Hermtian random matrices.

The study of the quenched SFF is particularly relevant if one aims to study glassy systems and in the future it would be interesting to investigate the properties of the zeros of the partition function of similar models \cite{obuchi2012partition} which are expected to play a role in the fluctuations of this quantity. It will also be interesting to explore the ``quenched versions" of survival probability that exhibits the interesting feature of a correlation hole~\cite{CH1986,TH2017,LS2018}. Understanding spikes in quenched $\sigma$FF built out of singular values of non-Hermitian systems is an interesting open problem~\cite{sig1,sig2,sig3}.

\section*{Acknowledgments}

We thank G. Bunin for collaborations on a related project. 
This research was supported in part by the International Centre for Theoretical Sciences (ICTS) for the program - Indo-French workshop on Classical and quantum dynamics in out of equilibrium systems  (code: ICTS/ifwcqm2024/12).
This work was supported by the French government through the France 2030 program (PhOM – Graduate School of Physics), under reference ANR-11-IDEX-0003 (Project Mascotte, L. Foini). M. K. acknowledges the support of the Department of Atomic Energy, Government of India, under project no. RTI4001.

\appendix
\setcounter{figure}{0}
\renewcommand{\thefigure}{A\arabic{figure}}

\section{Non hermitian random matrices}\label{App_RMT}
\label{Sec_App}

\begin{figure}
    \centering
    \includegraphics[width=8.25 cm]{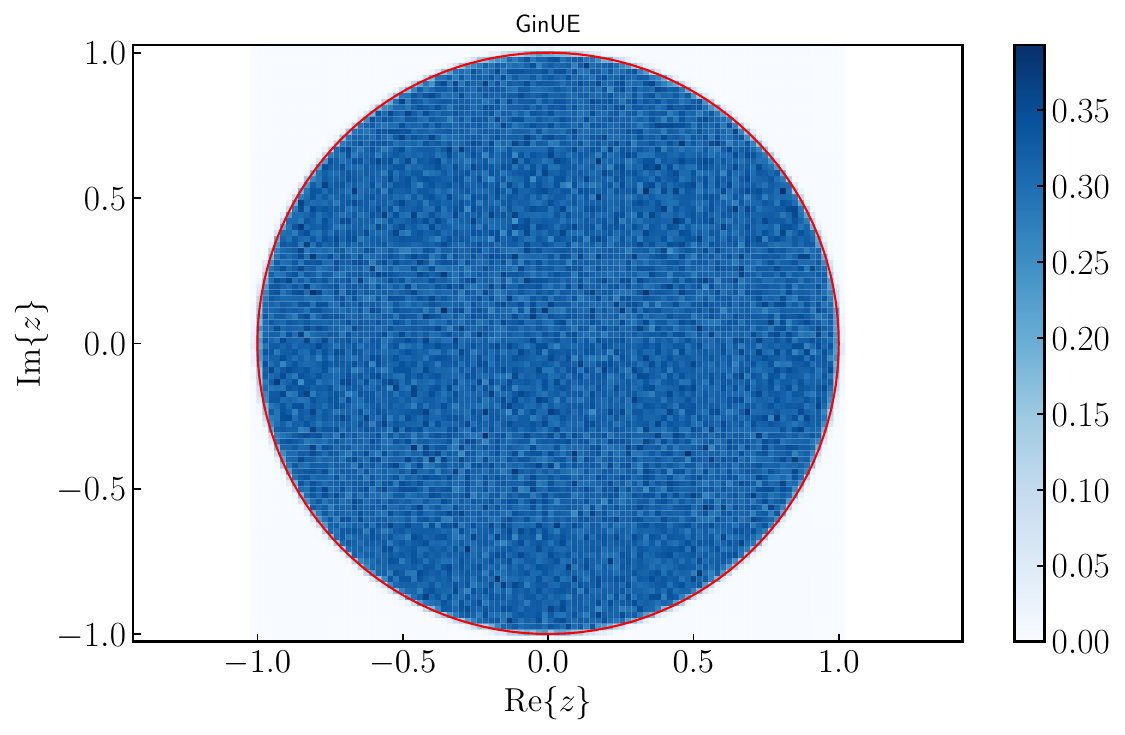}
    
    \caption{Eigenvalue distribution on the complex plane for the Ginibre Unitary Ensemble with $\mathcal{D}=3000$. The red curve is the unit circle.} 
    \label{fig:GinUEPDFfig}
\end{figure}

\begin{figure}
    \centering
    \includegraphics[width=8.25 cm]{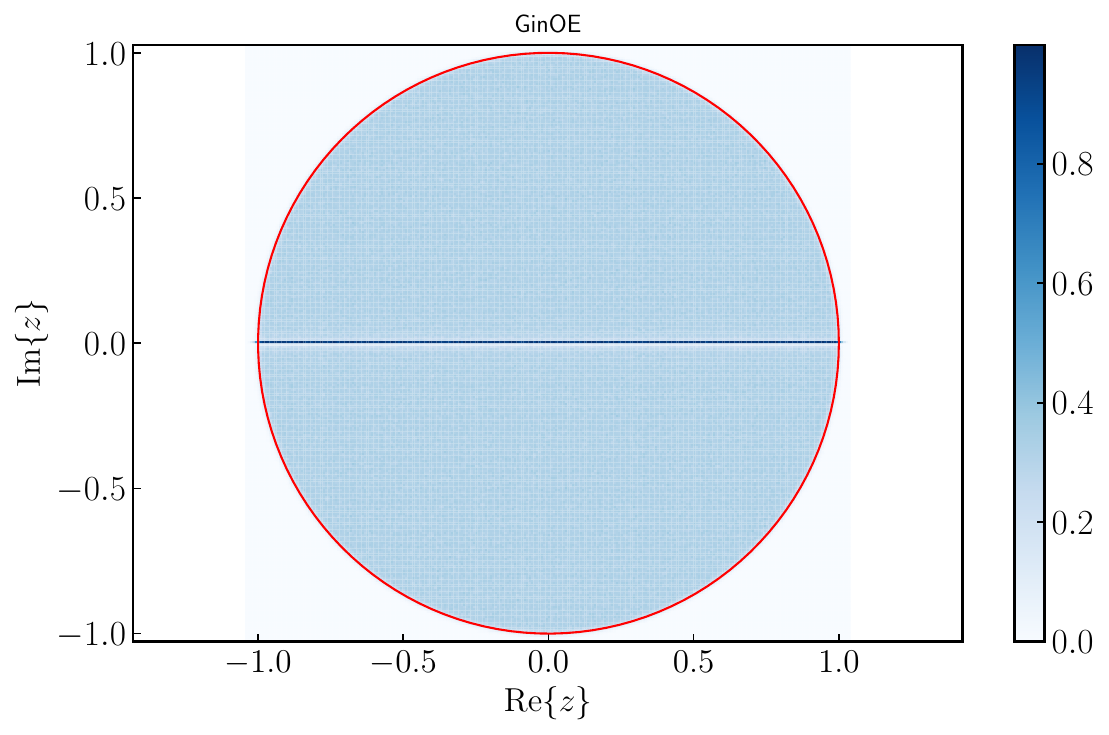}
    
    \caption{Eigenvalue distribution on the complex plane for the Ginibre Orthogonal Ensemble with $\mathcal{D}=3432$. The noticeable difference with the GinUE distribution of eigenvalues is the existence of a separate density on the real line. Moreover, complex eigenvalues come in conjugate pairs.
    }
    \label{fig:GinOEPDF}
\end{figure}

\begin{figure}[H]
    \centering
    \includegraphics[width=8.25cm]{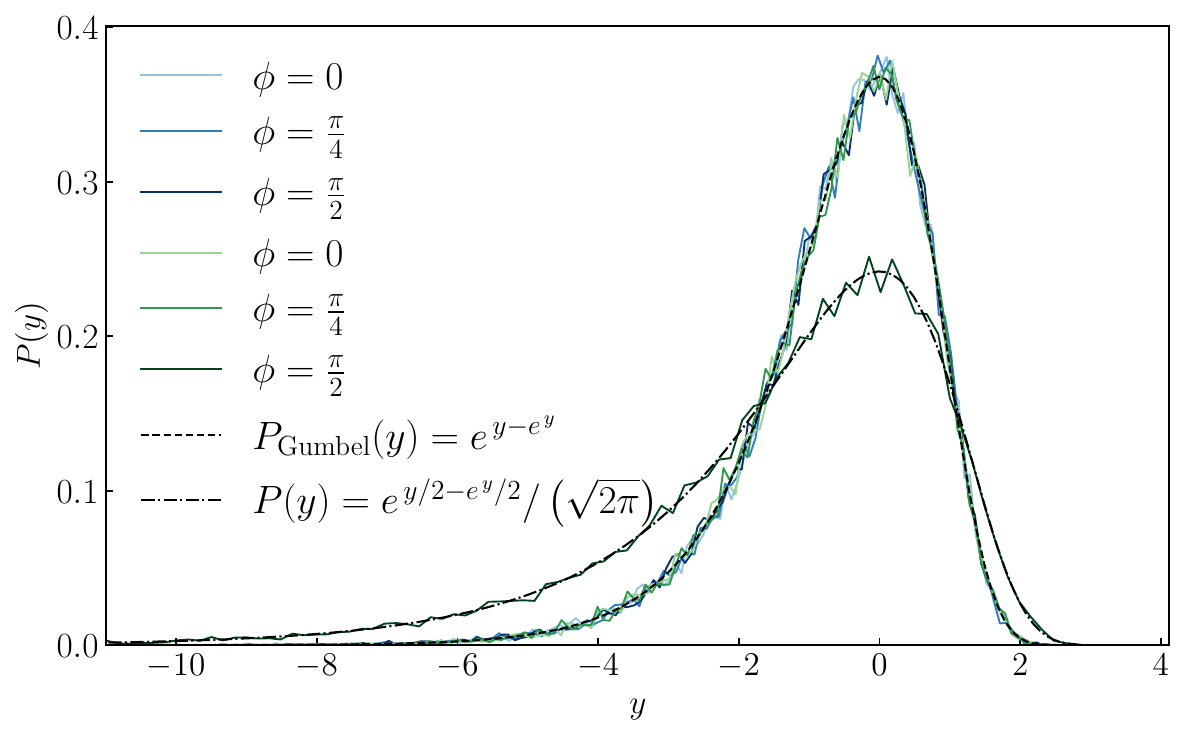}%
    \caption{Distribution of the variable $Y$ for a multitude of cases. We consider the GinUE and GinOE for $\phi=0,{\pi}/{4}$, and ${\pi}/{2}$ but note that we get identical results with $\phi=\pi/4$ for any angle between. The blue lines correspond to the empirical distributions obtained using as samples different instances of GinUE matrices whose $\mathrm{DSFF}$ is calculated at some time deep in the plateau. Likewise, the green lines correspond to the GinOE where we see a modified result for the case $\phi=\pi/2$. In both cases, we use $\mathcal{D}=200$.
    }
    \label{fig:Gumbel_ALL}
\end{figure}

In this appendix, we show results for complex non-hermitian random matrices and we start with a brief refresher on the Ginibre Unitary and Orthogonal Ensembles \cite{haake2018quantum,mehta2004random,Ginibre:1965zz,byun2023progressstudyginibreensembles,byun2023GinOEprogressstudyginibreensembles,tao2009randommatricesuniversalityesds,khoruzhenko2009nonhermitianrandommatrixensembles,PhysRevX.9.041015,PhysRevResearch.2.023286,PhysRevResearch.4.043196,Bernard_2002,Bordenave_2012,Tao_2015}. Each is comprised of matrices whose elements are independently distributed as $\sim\mathcal{N}(0,\sigma^2)$ complex or real numbers respectively. The joint PDF for matrices $A$ belonging to these ensembles in terms of their elements reads
\begin{equation}
    \label{eq:PDFelem}
    P(A)\propto e^{-\beta A^{\dagger}A/2},
\end{equation}
where $\beta=1,2$ for the GinOE and GinUE respectively. 
In the unitary case all the correlation functions can be determined through the kernel (Determinantal Point Process) 
\begin{equation}
    \label{eq:kernelGinUE}
    \mathcal{K}(z_1,z_2)=\frac{\mathcal{D}}{\pi}e^{-\mathcal{D}\frac{(|z_1|^2+|z_2|^2)}{2}}\sum_{n=0}^{\mathcal{D}-1}\frac{(\mathcal{D}z_1z_2^*)^n}{n!},\
\end{equation}
where $\mathcal{D}$ is the matrix dimension. In the asymptotic limit, the (properly rescaled) eigenvalues are uniformly distributed over the complex unit disk, i.e. $K(z,z)=\langle\rho(z)\rangle=1/\pi,|z|\leq1$\footnote{$\rho (z)=\sum_i\delta (x-x_i)\delta (y-y_i)$ is the spectral density for the rescaled $z\rightarrow z/\sqrt{\mathcal{D}}= x+iy$ eigenvalues with $\int dz\rho (z)=\mathcal{D}$.}, as dictated by the circular law, the Wigner semi-circle equivalent for non-hermitian random matrix ensembles with iid entries, and shown in Fig.~\ref{fig:GinUEPDFfig}. Their joint PDF has a form similar to that of the GUE:
\begin{equation}
    \label{eq:GinUEPDF}
    P_{\mathcal{D}}(z_1,\ldots,z_{\mathcal{D}})= \frac{1}{C_{\mathcal{D}}}e^{-\sum_{m=1}^{\mathcal{D}}|z_m|^2}\prod_{1\leq m<n\leq\mathcal{D}}|z_m-z_n|^2,
\end{equation}
with $C_{\mathcal{D}}=\pi^{\mathcal{D}}\prod_{j=1}^{\mathcal{D}}j!$.
When the matrix elements become real, a random (sub-extensive) number of the eigenvalues are real, and the rest come in complex conjugate pairs and their joint PDF is notably more complex. The rescaled spectral densities $\rho(x)$ and $\rho(z)$ are considered independently, and it can be nevertheless shown for both that in the limit $\mathcal{D}\to\infty$, they are uniform in the line $[-1,1]$ and unit disk (away from the edges), as can be seen with a numerical example in Fig.~\ref{fig:GinOEPDF}.
Fig.~\ref{fig:Gumbel_ALL} confirms the general validity of the Gaussianity assumption for the $\mathrm{DSFF}$ through the empirical distribution of the variable $Y$ for both the GinUE and GinOE. The existence of real eigenvalues modifies the expected distribution in a way that is described in Sec~\ref{subsec:Non-Herm-Many-Body}.

\bibliography{biblio}

\end{document}